\documentclass[a4paper,fleqn]{cas-sc}

\usepackage[authoryear]{natbib}
\usepackage{caption}
\usepackage{subcaption}
\usepackage[utf8]{inputenc}
\usepackage{threeparttable}

\def\tsc#1{\csdef{#1}{\textsc{\lowercase{#1}}\xspace}}
\tsc{WGM}
\tsc{QE}
\tsc{EP}
\tsc{PMS}
\tsc{BEC}
\tsc{DE}

\begin{document}
\let\WriteBookmarks\relax
\def\floatpagepagefraction{1}
\def\textpagefraction{.001}

\shorttitle{Evaluating raw waveforms with deep learning frameworks for speech emotion recognition}

\shortauthors{Kilimci et~al.}

\title [mode = title]{Evaluating raw waveforms with deep learning frameworks for speech emotion recognition}                      




\author[a]{Zeynep Hilal Kilimci\corref{mycorrespondingauthor}}
\ead{zeynep.kilimci@kocaeli.edu.tr}
\cortext[mycorrespondingauthor]{Corresponding author: Ayhan Küçükmanisa}

\credit{Conceptualization of this study, Methodology, Writing - Original Draft, Supervision}
\author[b]{Ulkü Bayraktar}
\ead{ulkubayraktar@gmail.com}
\credit{Software, Formal analysis, Investigation}
\author[b]{Ayhan Küçükmanisa}
\ead{ayhan.kucukmanisa@kocaeli.edu.tr}
\credit{Conceptualization of this study, Methodology, Writing - Original Draft, Supervision}

\address[a]{Department of Information Systems Engineering, Kocaeli University, 41001, Kocaeli, Turkey}
\address[b]{Department of Electronics and Communication Engineering, Kocaeli University, 41001, Kocaeli, Turkey}

\begin{abstract}
Speech emotion recognition is a challenging task in speech processing field. For this reason, feature extraction process has a crucial importance to demonstrate and process the speech signals. In this work, we represent a model, which feeds raw audio files directly into the deep neural networks without any feature extraction stage for the recognition of emotions utilizing six different data sets, namely, The Berlin Database of Emotional Speech (EMO-DB), Ryerson Audio-Visual Database of Emotional Speech and Song (RAVDESS), Toronto Emotional Speech Database (TESS), Crowd-sourced Emotional Multimodal Actors (CREMA), Surrey Audio-Visual Expressed Emotion (SAVEE), and TESS+RAVDESS. To demonstrate the contribution of proposed model, the performance of traditional feature extraction techniques namely, mel-scale spectogram, mel-frequency cepstral coefficients, are blended with machine learning algorithms, ensemble learning methods, deep and hybrid deep learning techniques. Support vector machine, decision tree, naive Bayes, random forests models are evaluated as machine learning algorithms while majority voting and stacking methods are assessed as ensemble learning techniques. Moreover, convolutional neural networks, long short-term memory networks, and hybrid CNN-LSTM model are evaluated as deep learning techniques and compared with machine learning and ensemble learning methods. To demonstrate the effectiveness of proposed model, the comparison with state-of-the-art studies are carried out. Based on the experiment results, CNN model excels existent approaches with 95.86\% of accuracy for TESS+RAVDESS data set using raw audio files, thence determining the new state-of-the-art. The proposed model performs 90.34\% of accuracy for EMO-DB with CNN model, 90.42\% of accuracy for RAVDESS with CNN model, 99.48\% of accuracy for TESS with LSTM model, 69.72\% of accuracy for CREMA with CNN model, 85.76\% of accuracy for SAVEE with CNN model in speaker-independent audio categorization problems.

\end{abstract}




\maketitle

\section{Introduction}

Emotions are complex psychophysiological changes resulting from the interaction of individual moods with biochemical and environmental influences. This change can be expressed in different ways such as speech, facial expression, body motions, and brain signs by emphasizing emotions like anger, sadness, happiness, fear, excitement, surprise, etc. Speech is a complicated sign that includes significant information related to the content of message and features of speaker (gender, emotion, language, accent, etc.). That is why it has been explored by many disciplines and art forms. 

Speech emotion recognition (SER) is a common research field in the last decades. SER is utilized in different application areas such as human-machine interaction, education, management of multimedia contents, entertainment, automobile industry, text to speech conversion, medical diagnosis \citep{ingale2012speech}. 

Emotion recognition from speech signal is reasonably hard since the styles of speaking (i.e. pronunciation), speech rates of the speakers is totally diverse from individual to individual and it modifies from location to location (i.e. distinct for native speakers and non-native speakers) \citep{ingale2012speech}. Thence, it is more significant to pick up specific attributes of speech which are not influenced by the territory, culture, speaking genre of the talker. Various characteristics such as spectral, prosodic, and acoustic  are employed by extracting features from speech signal for emotion recognition task in computer science \citep{selvaraj2016human}. Then, the procedure is proceeded by classifiers to determine the emotion of speech.

Deep learning-based models are commonly employed by the researchers because of providing better predictions and results compared with traditional machine learning algorithms in different domains such as face recognition, voice recognition, image recognition \citet{mittal2018real}, \citet{bae2016voice},  \citet{he2016deep}. The usage of deep learning architectures facilitates automatic feature selection process unlike gathering hand-crafted features. Lately, diverse deep learning-based models are also utilized for speech emotion recognition tasks in the state-of-the-art-studies. The literature works generally focus on to discover important attributes of speech signal using deep learning models \citet{trigeorgis2016adieu} or demonstrate the performance of deep learning models on a specific feature extraction technique \citet{han2014speech}. In this work, we obtain deep features from raw sound data and feeding them into different deep learning algorithms for speech emotion recognition task instead of employing conventional feature extraction techniques such as MFCC, MEL.

In this work, evaluating raw waveforms is proposed without applying any feature extraction stage using traditional machine learning algorithms, ensemble learning approaches, and deep learning architectures for speech emotion recognition task. For this purpose, convolutional neural networks, long short-term memory networks, and hybrid CNN-LSTM models are evaluated deep learning techniques while support vector machine, decision tree, naive Bayes, random forests, majority voting, stacking models are evaluated as machine learning and ensemble algorithms. To demonstrate the contribution of the proposed framework, the performance of traditional feature extraction techniques and the results of state-of-the-art studies are compared with the proposed model on six different dataset. The utilization of raw audio files instead of employing feature extraction process for speech emotion recognition task shows remarkable results when compared to the literature studies.

The remaining of the paper is organized as follows: Section \ref{sec2} presents literature review. Materials and methods used in the study are given in Section \ref{sec3}. Data acquisition and proposed framework are demonstrated in Section \ref{sec4}. In Section \ref{sec5}, experiment results and conclusions are represented.

\section{Literature Review}\label{sec2}

This section gives a brief summary of literature works for speech emotion recognition.

\citep{issa2020speech} introduce convolutional neural network (CNN) architecture is introduced for speech emotion recognition task. Mel-frequency Cepstral Coefficients (MFCCs), Mel-scaled spectrogram, Chromagram, Spectral contrast feature, and Tonnetz representation are evaluated at the stage of feature extraction. After that, extracted features from sound files are fed into CNN to show the effectiveness of feature extraction models in RAVDESS, EMO-DB, and IEMOCAP data sets. The proposed model exhibits the best classification performance in EMO-DB data set with 86.1{\%} of accuracy. \citep{sajjad2020clustering} propose a new clustering based approach with the help of radial-based function network for SER. Determined key sequences are fed into Bidirectional long short-term memory network to obtain final state of the emotion. The proposed approach is assessed over IEMOCAP, EMO-DB, and RAVDESS data sets. Experiment results show that the proposed approach represents remarkable results in terms of classification accuracy when compared to the state-of-the-art studies. 

\citep{kwon2019cnn} presents a CNN-based framework for speech emotion recognition. To show the effectiveness of the model, IEMOCAP and RAVDESS data sets are evaluated. The authors report that the proposed model enhance the classification accuracy by 7.85{\%} for IEMOCAP and 4.5{\%} for RAVDESS. \cite{zhao2019speech} evaluate 1D and 2D CNN-LSTM networks to recognize speech emotion. The performance of hybrid deep learning model is compared with deep belief network and CNN architecture over two data sets namely, IEMOCAP and EMO-DB. They report that the performance of hybrid deep learning model for SER is rather competitive when compared to the conventional techniques. \citep{koduru2020feature} investigate the impact of feature extraction techniques to enhance the performance of speech emotion rate. For this purpose, Mel frequency cepstral coefficients, Discrete Wavelet Transform (DWT), pitch, energy and Zero crossing rate (ZCR) models are employed at the stage of feature extraction. To show the efficieny of feature extraction techniques, support vector machine, decision tree (DT) and LDA models are evaluated. The utilization of DT performs the best classification accuracy with nearly 85{\%}. 

\citep{kwon2021mlt} propose a multi-learning strategy by providing end-to-end real time model for SER. The proposed dilated CNN (DCNN) model is evaluated on two benchmark data sets namely, IEMOCAP and EMO-DB. Authors report that the usage DCNN model exhibits significant accuracy results with 73{\%} for IEMOCAP and 90 {\%} for EMO-DB. \citep{zehra2021cross} propose an ensemble-based framework for cross corpus multi-lingual recognition of speech emotion. The performance of an ensemble model, majority voting, is compared with conventional machine learning techniques. The utilization of ensemble model ensures enhancement in classification accuracy nearly 13{\%} for Urdu data set, roughly 8{\%} for German data set, 11{\%} for Italian data set, and 5{\%} for English data set. \citep{anvarjon2020deep} concentrate on a lightweight CNN approach for speech emotion recognition task. To show the efficiency of proposed model, experiments are carried on IEMOCAP, and EMO-DB data sets. Experiment results indicate that a lightweight CNN model is capable to recognize emotion of speech with 77.01 {\%} of accuracy for IEMOCAP data set, and 92.02{\%} of accuracy for EMO-DB data set. \citep{ozseven2019novel} presents a novel statistical feature selection technique by taking into consideration average of each featue in the features set for SER. Recognition performance of the model is compared on EMO-DB, eNTERFACE05, EMOVO and SAVEE data sets by using SVM, MLP, and k-NN classifiers. Except eNTERFACE05 data set, SVM model outperforms other machine learning algorithms in terms of classification accuracy in all data sets. 

\citep{sun2019speech} propose DNN-decision tree SVM model by calculating the confusion degree of emotion with decision tree SVM and training with DNN architecture. To demonstrate the effectiveness of the proposed model, experiments are carried on Chinese Academy of Sciences Emotional data set. The proposed model achieves remarkable experiment results when compared to conventional SVM and  DNN-SVM technique by ensuring nearly 6{\%} and roughly 3{\%} enhancement in the success of recognition rate, respectively.\citet{cai2021speech} focus on the multitask learning approach for speech emotion recognition by ensuring speech-to-text recognition and emotion categorization, simultaneously. The efficiency of the model is demonstrated on the IEMOCAP data set by achieving nearly 78{\%} of accuracy. \citep{chen2020two} present two-layer fuzzy multiple ensemble framework using fuzzy C-means algorithm and random forest model for SER. The proposed framework is capable to recognize emotions on CASIA and EMO-DB data sets by improving recognition accuracy when compared to back propagation and random forest models. \citep{farooq2020impact} concentrate on the effect of feature selection model utilizing deep convolutional neural network (DCNN) for SER. After extracting features from pretrained models, the most discriminatory features are determined by a correlation-based feature selection technique. At the classification stage, support vector machines, random forests, the k-nearest neighbors algorithm, and neural network classifier are employed on EMO-DB, SAVEE, IEMOCAP, and RAVDESS data sets. The model achieves 95.10{\%} for Emo-DB, 82.10{\%} for SAVEE, 83.80{\%} for IEMOCAP, and 81.30{\%} for RAVDESS in terms of classification accuracies.

\citep{zhang2021learning} propose a novel deep multimodal model for spontaneous SER. The proposed model is based on the combination of three different audio inputs by feeding them into multi-CNN fusion model. The combination strategy performs promising classification results when compared with the sate-of-the-art results. \citep{tuncer2021automated} focus on SER by employing twine shuffle pattern and iterative neighborhood component analysis methods. The proposed model is based on feature generation and selection stages utilizing shuffle box and iterative neighborhood component analysis methodologies, respectively. To demonstrate the efficiency of the model, the experiments are carried out on EMO-DB, SAVEE, RAVDESS, EMOVO data sets. Proposed model achieves 87.43{\%} for RAVDESS, 90.09{\%} for EMO-DB, 84.79{\%} for SAVEE, and 79.08{\%} for EMOVO in terms of classification accuracy. \citep{lu2022domain} present domain invariant feature learning (DIFL) models to address speaker-independent speech emotion recognition. The experiments are performed on EMO-DB, eNTERFACE, and CASIA data set to demonstrate the contribution of the proposed model. Experiment results indicate that the utilization of proposed model exhibits remarkable results compared to the literature studies.

To the best of our knowledge, our study is the first attempt to process raw audio files by blending them with machine learning and deep learning methods for the task of speech emotion recognition and differs from the aforementioned literature studies in this aspect.

\section{Datasets and methodology}\label{sec3}

\begin{figure}
\begin{center}

    \begin{subfigure}{0.8\textwidth}
    \includegraphics[width=\linewidth]{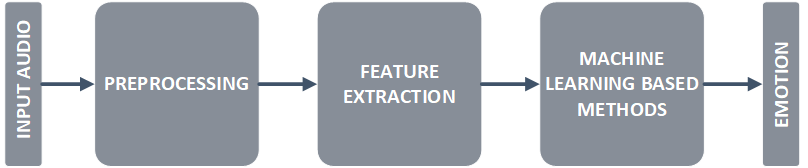}
    \caption{}
    \end{subfigure}\hfil 
    \begin{subfigure}{0.8\textwidth}
    \includegraphics[width=\linewidth]{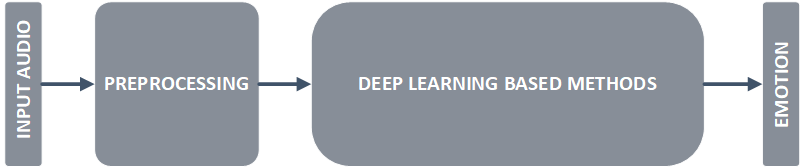}
    \caption{}
    \end{subfigure}\hfil 
    
\caption{The block diagram of proposed methodology (a) ML based methodology (b) DL based methodology}
\label{blockDiagram}
\end{center}
\end{figure}

In this section, datasets used in the study, and proposed methodology are presented. Six different publicly available and widely applied datasets, The Berlin Database of Emotional Speech (EMO-DB), The Ryerson Audio-Visual Database of Emotional Speech and Song (RAVDESS), Toronto Emotional Speech Database (TESS), Crowd-sourced Emotional Multimodal Actors dataset (CREMA), Surrey Audio-Visual Expressed Emotion (SAVEE), and TESS+RAVDESS are employed in the experiments. After that, methodology is introduced with feature extraction stage, models, and the proposed framework.  The proposed speech emotion recognition methodology is demonstrated in Figure \ref{blockDiagram}.

\subsection{Datasets}
\begin{figure}
\begin{center}
    \begin{subfigure}{0.5\textwidth}
    \includegraphics[width=\linewidth]{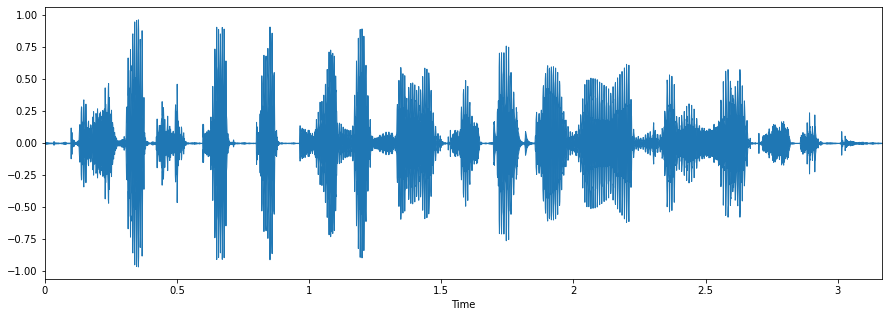}
    \caption{}
    \end{subfigure}\hfil 
    \begin{subfigure}{0.5\textwidth}
    \includegraphics[width=\linewidth]{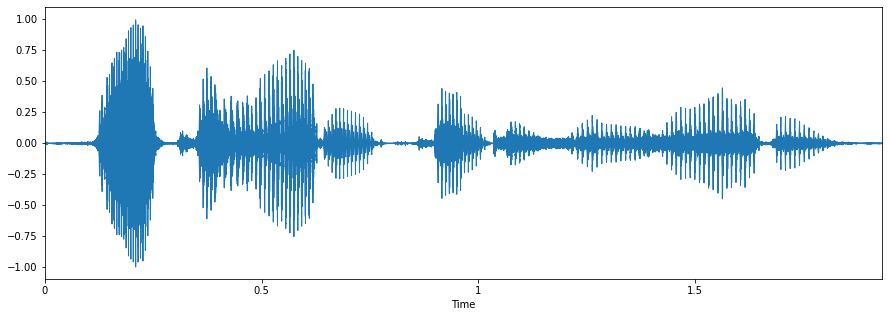}
    \caption{}
    \end{subfigure}\hfil 
    \begin{subfigure}{0.5\textwidth}
    \includegraphics[width=\linewidth]{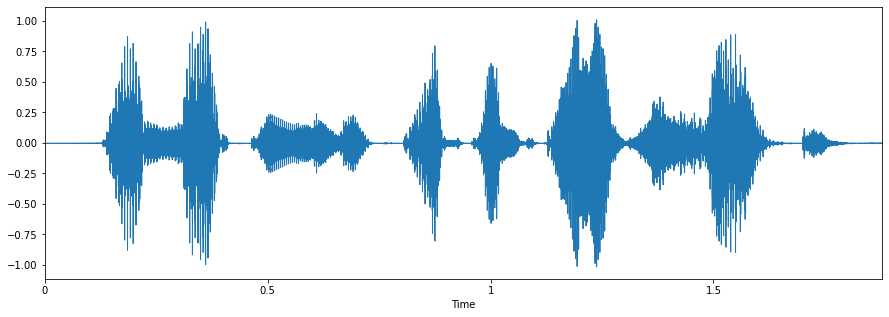}
    \caption{}
    \end{subfigure}\hfil 
    \begin{subfigure}{0.5\textwidth}
    \includegraphics[width=\linewidth]{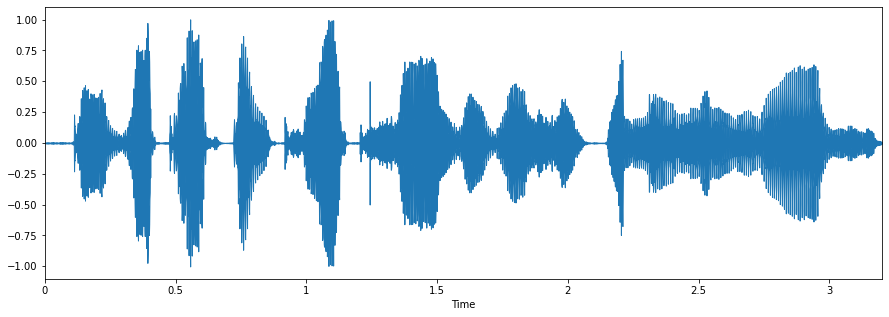}
    \caption{}
    \end{subfigure}
    \begin{subfigure}{0.49\textwidth}
    \includegraphics[width=\linewidth]{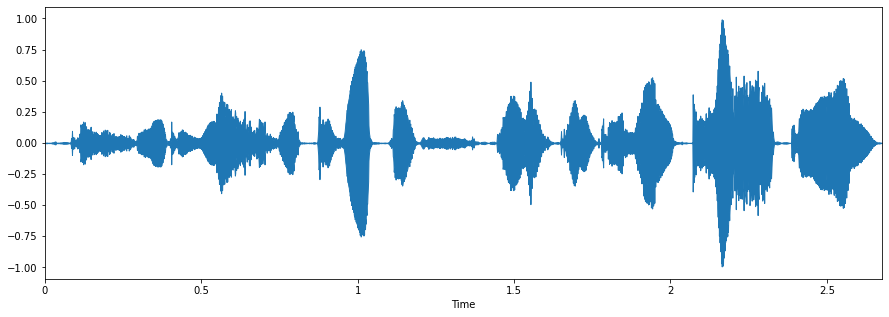}
    \caption{}
    \end{subfigure}
    \begin{subfigure}{0.49\textwidth}
    \includegraphics[width=\linewidth]{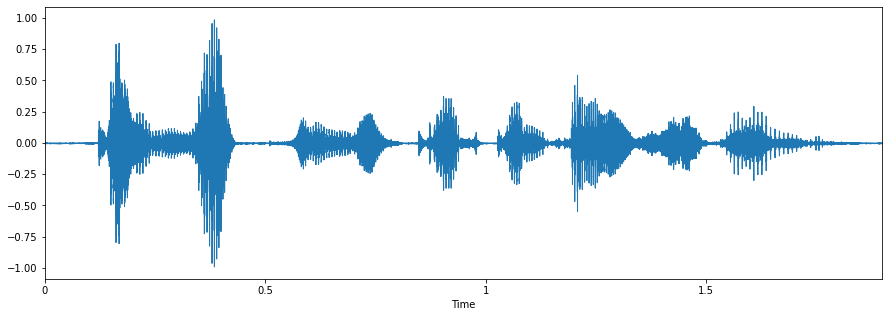}
    \caption{}
    \end{subfigure}
    \begin{subfigure}{0.49\textwidth}
    \includegraphics[width=\linewidth]{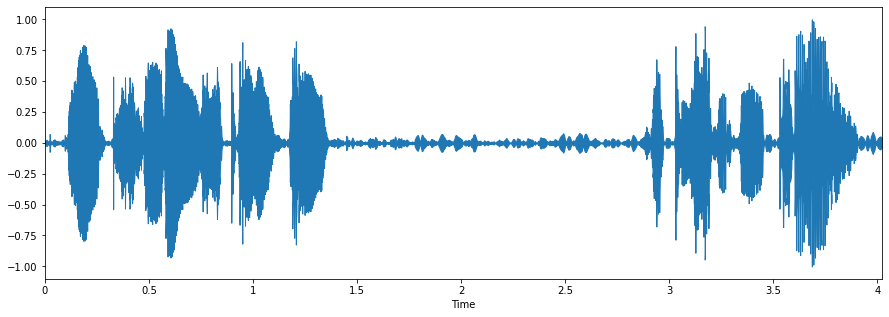}
    \caption{}
    \end{subfigure}
\caption{The wave forms of seven emotions from EMO-DB dataset (a) Neutral (b) Bored (c) Angry (d) Disgust (e) Fear (f) Happy (g) Sad.}
\label{Figure1}
\end{center}
\end{figure}

Six different data sets (EMO-DB, RAVDESS, TESS, CREMA, SAVEE, TESS+RAVDESS) are utilized to enhance the generalization capacity of the consequences acquired in the proposed work. EMO-DB dataset \citet{burkhardt2005database} is composed of definition of different feelings (neutral, happiness, anger, disgust, sadness, boredom, and fear/anxiety) by ten different actors whom of 5 is female. The dataset includes 535 words in German and each of audio file has 16 kHz sampling frequency with 16 bit quantization. The waveform of each emotion is represented in Figure \ref{Figure1}. RAVDESS dataset \citep{livingstone2018ryerson}  contains English sentences voiced by 12 male and 12 female actors. RAVDESS is composed of 1,440 utterances with eight different emotion categories, namely, surprised, angry, happy, bored, sad, fearful, disgust, and neutral.

TESS \cite{SP2/E8H2MF_2020} comprises of audio records of 2 female speakers pronouncing English sentences. TESS contains 2,800 utterances of  with anger, disgust, neutral, fear, happiness, sadness, bored, surprise emotional categories. CREMA \citep{cao2014crema} includes 7,442 original clips generated by 43 female and 48 male actors from various ethnicities such as Hispanic, Asian, African, American. Specified 12 sentence are vocalized capturing six different emotions. These are sad, happy, disgust, neutral, anger, fear. SAVEE dataset \citet{haq2008audio} covers 1,680 utterances vocalized by 14 male actors in seven different emotions. These are surprise, anger, happiness, disgust, neutral, fear, sadness. The sentences recorded by actors are picked up from the TIMIT Acoustic-Phonetic Continuous Speech corpus. The combination of TESS and RAVDESS data sets is called TESS+RAVDESS in this study. Because the emotional categories are common, there is no problem to consolidate them. The dataset consists of 4,240 utterances voiced by 14 female and 12 male actors. The distribution of the datasets used in the study is given in Table \ref{table:Table1}.

\begin{table}[!ht]
\centering
\caption{Distribution of emotion categories for each dataset}
\begin{tabular}{cccccccccc}
\hline\noalign{\smallskip}
\textbf{Datasets} & \textbf{NR} & \textbf{HP} & \textbf{AG} & \textbf{SR} & \textbf{SD} & \textbf{DG} & \textbf{FR} & \textbf{BR} & \textbf{Total} \\
\noalign{\smallskip}\hline\noalign{\smallskip}
\textbf{EMO-DB} & 79 & 71 & 127 & - & 62 & 46 & 81 & 69 &535 \\
\textbf{RAVDESS}  & 288 & 192 & 192 & 192 & 192 & 192 & - & 192 & 1440 \\
\textbf{TESS}  & 400 & 400 & 400 & 400 & 400 & 400 & - & 400 & 2800\\
\textbf{CREMA}  & 1271 & 1271 & 1271 & - & 1271 & 1271 & - & 1271 & 7626 \\
\textbf{SAVEE}  & 120 & 60 & 60 &  60 & 60 & 60 & - & 60 & 480 \\
\textbf{TESS+RAVDESS}  & 688 & 592 & 592 & 592 & 592 & 592 & - & 592 & 4240\\
\noalign{\smallskip}\hline \\
\end{tabular}
\begin{tablenotes}
      \small
      \item NR: Neutral, HP: Happy, AG: Angry, SR: Surprised, SD: Sadness, DG: Disgust, FR: Fear, BR: Bored
\end{tablenotes}
\label{table:Table1}
\end{table}

\subsection{Feature extraction and pre-processing}
Feature extraction stage plays an importance role at specifying the performance of any learning methodology. Eligible selection of feature could enable to a better trained method, while inconvenient features would crucially disrupt the training procedure \citet{trigeorgis2016adieu}. In this work, we mainly focus on to detect the speech emotion from raw audio files without using hand-crafted features. The feature extraction stage is automatically carried out in the deep learning architectures by processing raw audio files, directly. To show the effectiveness of proposed model, the performance of the system is compared with the conventional feature extraction techniques, namely, Mel-scale Spectogram and Mel-Frequency Cepstral coefficients (MFCC).

Mel-scaled spectrogram is widely applied in sound classification and speech emotion recognition tasks \citep{stevens1937scale}. The features obtained with Mel-scaled spectrogram makes possible to imitate the sound frequency of human in a specific rank. It is known that Mel-scaled spectrogram performs well recognition and, pursing timbre fluctuations in an audio file. On the other hand, it inclines to be weak when a distinguishable representation of pitch classes and harmony are considered \cite{beigi2011fundamentals}. Log-mel spectrogram feature set comprises in Mel-scaled spectrograms demonstrating emotion states. At the pre-processing stage, noise reduction, windowing, framing are carried out to the speech signal. 1024 size of short time Fourier transform and 0.025 s and 0.010 s overlapped window size with hamming window are employed in the experiments. In addition, 128 equal-width log-energies are used. Finally, 168 features are acquired as a result of Mel-spectogram processes.

The MFCC \cite{Dave2013} feature extraction technique is considered to be the closest feature extraction technique to the human auditory system. Initially, in this procedure, the original signal is translated from the time domain to the frequency domain using the discrete Fourier transform (DFT). For this conversion, the power spectrum is employed. For the purpose of reducing frequency distortion brought on by segmentation prior to DFT, hamming window is utilized. The frequency is then wrapped from the hertz scale to the mel scale using a filter bank. In conclusion, the logarithm of the Mel scale power spectrum's feature vectors are extracted using discrete cosine transformation (DCT) \cite{Sunitha2015}. At the pre-processing stage, as same as with Mel-spectogram noise reduction, windowing, framing are carried out to the speech signal. 40 features are acquired as a result of MFCC processes. The size of the MFCC features varies depending on variations in the input audio size. Because of this, the first 2.5 seconds of each sound are used for feature extraction.

Training acoustic models straight from the raw wave-form data is challenging task in speech recognition field. Traditional deep neural network-based acoustic methods is based on processing hand-crafted input features. In this work, we propose deep neural network-based acoustic method which fed with raw multichannel waveforms as input by inspiring studies \citet{dai2017very}, \citet{hoshen2015speech}, \citep{palaz2013estimating}, \citet{tuske2014acoustic}, \citet{jaitly2011learning} without performing feature extraction stage by constructing high-level representative features. As a pre-processing step, raw sounds are first normalized to mean 0 and variance 1. If the length of the audio data is more than the specified upper limit size (6 sn), it is clipped. Otherwise if length is lower than threshold, input array is padded with "0" values.

\subsection{Machine learning based methods}
Machine learning which is a subfield of artificial intelligence, is a way of teaching computers how to do things that people can do naturally, like learning to recognize patterns in data. There are different machine learning algorithms, each of which is good at solving different types of problems. In this work, some of popular machine learning algorithms and their ensemble versions are used to solve speech emotion recognition problem.

\textbf{SVM} \citep{Cristianini2000} is a machine learning algorithm used for classification (determining whether objects belong to a certain category) and regression (in predicting future values). It is particularly useful for classification problems in which the objects in the data set can be neatly divided into groups, or classes. The SVM algorithm helps us to divide a space into categories so that we can easily put new data points into the right category in the future. This is done by creating a decision boundary that separates the different categories. The Support Vector Machine (SVM) selects the points along the axes that are most helpful in creating a hyperplane. These points are called "support vectors," and this is the reason algorithm is called as a Support Vector Machine.

\textbf{k-NN} \citep{Mucherino2019} algorithm assumes that things that are similar tend to be near each other. It is the simplest machine learning algorithm. KNN is basically based on the premise that the class values of nearest samples will be similar. In this evaluation two values are used. Distance: The distance of the k nearest sample of the sample whose class value is to be calculated. K (neighbor count): It is the number of nearest neighbors which this calculation will be made.

\textbf {Decision tree} \citep{Rokach2015} is a algorithm that can help model situations and make decisions based on dividing the data set into smaller subgroups that are within the framework of certain rules (decision rules). It has a tree-like structure, with branches representing different possible outcomes. Trees are also useful for modeling resource costs and possible outcomes for decision making. The tree structure contains some units. Internal nodes representing the tests or attributes of each stage. Each branch indicate an attribute result. At final, the path from leaf to root represents rules of classification.

\textbf {Naïve Bayes} \citep{Webb2010} classification is a method used to estimate the probability that a particular set of features belongs to a particular class. It uses the Bayes theorem to calculate the probability of each class, and then selects the class with the highest probability. This method is much faster than more complex methods, and is often used to quickly determine which class a particular set of features belongs to.

Ensemble methods are a way to combine the predictions of several different estimators to improve the accuracy and reliability of predictions. One of the most known and used ensemble methods is \textbf {majority voting}. In this approach, different models perform predictions and the prediction with the most votes is determined as the final decision.

Another popular ensemble approach is \textbf {stacking} \citep{Dzeroski2004}. This approach includes a two-level learning process, Base (Level 0) and Meta (Level 1). The base classifiers run in parallel and their estimations are combined into a metadata. Then these estimations become input into the meta classifier. Basically, stacking approach tries to figure out the best way to combine the input predictions to get a more accurate output prediction.

\subsection{Deep learning based methods}

In this work, 3 different deep learning approaches are used: conventional CNN, LSTM and CNN-LSTM. Deep learning is considered as the cutting edge of artificial intelligence today. It is basically inspired by the human learning system. With its hierarchical connections and its multi-layered structure, it enables the learning from the lowest level features to the highest level features. The values of the weights of the connections in the layers are the most important point of the learning function. Convolutional Neural Networks (CNN) is developed based on the specialization of neurons in the visual and perception system of humans. While 2-dimensional filters are used in 2-dimensional images, learning function is performed with 1-dimensional filters in 1-dimensional data such as sound or time series.

Recurrent Neural Networks (RNNs) are a type of Deep Learning structures used to predict the next step. RNNs use the output of the previous step as the input of the current step, whereas classical deep learning networks work independently of each other. As a result, the RNN ensures that each output it generates follows the previous step. As a result, it tries to store the results of the previous steps in its memory. However, they are successful in predicting short-term dependencies, they are not successful enough in long-term dependencies. Because of these fundamental RNN problems, later variants of Long Short-Term Memory (LSTM) \citep{Hochreiter1997} networks have been proposed.

LSTM is proposed as a solution to the short-term memory problem. It solves this problem using Cell State and various gates. Cell State is a line that carries meaningful information across cells, and the gates use the sigmoid activation function to squash the data between 0 to 1. If the values that the sigmoid activation function can have are taken into consideration, 0 means that the information will be forgotten, and 1 means that it will continue to be used as it is. LSTM has 3 gates as forget gate, input gate, and output gate. The Forget Gate determines amount of information in a memory will be forgotten or kept. The Input Gate updates the Cell State based on the results of the sigmoid process. The Output Gate decides what the next cell's input will be. 

\begin{figure}[htb!]
\begin{center}
\includegraphics[width=1\linewidth]{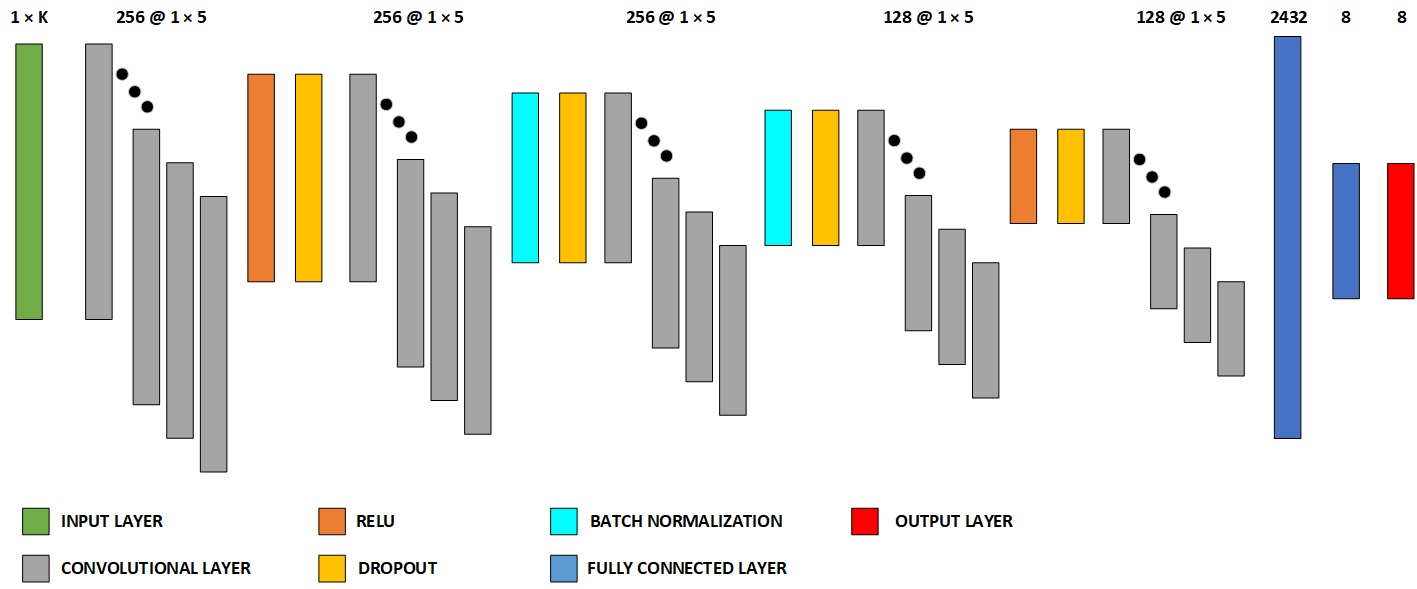}
\caption{CNN architecture}
\label{CNNModel}
\end{center}
\end{figure}

The proposed CNN model which is depicted in Figure \ref{CNNModel} consists of one-dimensional convolutional layers consolidated with activation, batch normalization and dropout layers. The first layer is consisted of 256 filters with the kernel size of 1 × 5 and stride 1. After that, the output is activated by using rectified linear unit (ReLU) and dropout layer with the ratio of 0.25 is performed. The second layer and the third layer is also constructed with 256 filters, and the same stride and kernel size are processed similarly preceding layer. In these 2 layers after convolution, batch normalization is carried out sending its output to the dropout layer with the ratio of 0.25. After that, convolution layers with 128 filters of size 1 × 5 is applied in fourth and fifth layers and before fifth convolutional layer ReLU activation and dropout layers is used. Then, 2 Fully Connected Layers (FCN) with 2432 and 8 are pursued after the last convolutional layer. Finally, output layer has 8 output for emotions and it uses Softmax function.

\begin{figure}[htb!]
\begin{center}
\includegraphics[width=1\linewidth]{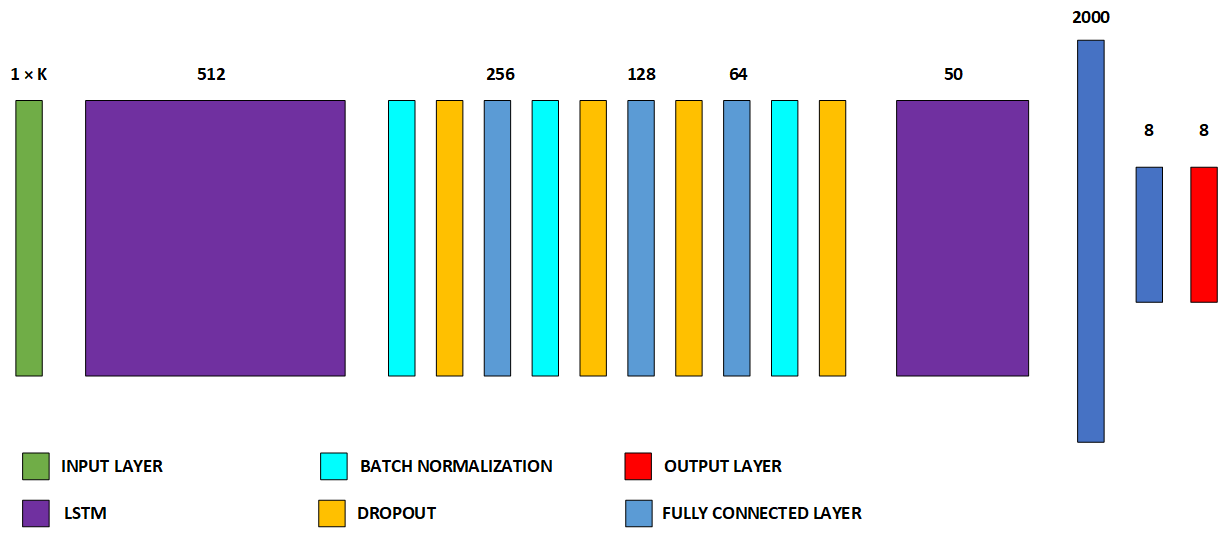}
\caption{LSTM architecture}
\label{LSTMModel}
\end{center}
\end{figure}

The proposed LSTM network used in this work shown in Figure \ref{LSTMModel}. First, LSTM block with 512 node is performed on input data. Then batch normalization, dropout layer with ratio of 0.25 and FCN with 256 size is applied, respectively. Next, batch normalization, dropout layer with ratio of 0.25 and FCN with 128 size is constructed. After that dropout and FCN with 64 is carried out. Then, batch normalization is carried out sending its output to the dropout layer with the ratio of 0.25. Afterwards, it is continued 2 Fully Connected Layers (FCN) with 2000 and 8 after the last LSTM with 50 nodes. Finally, output layer has 8 output for emotions and it uses Softmax function.

\begin{figure}[htb!]
\begin{center}
\includegraphics[width=1\linewidth]{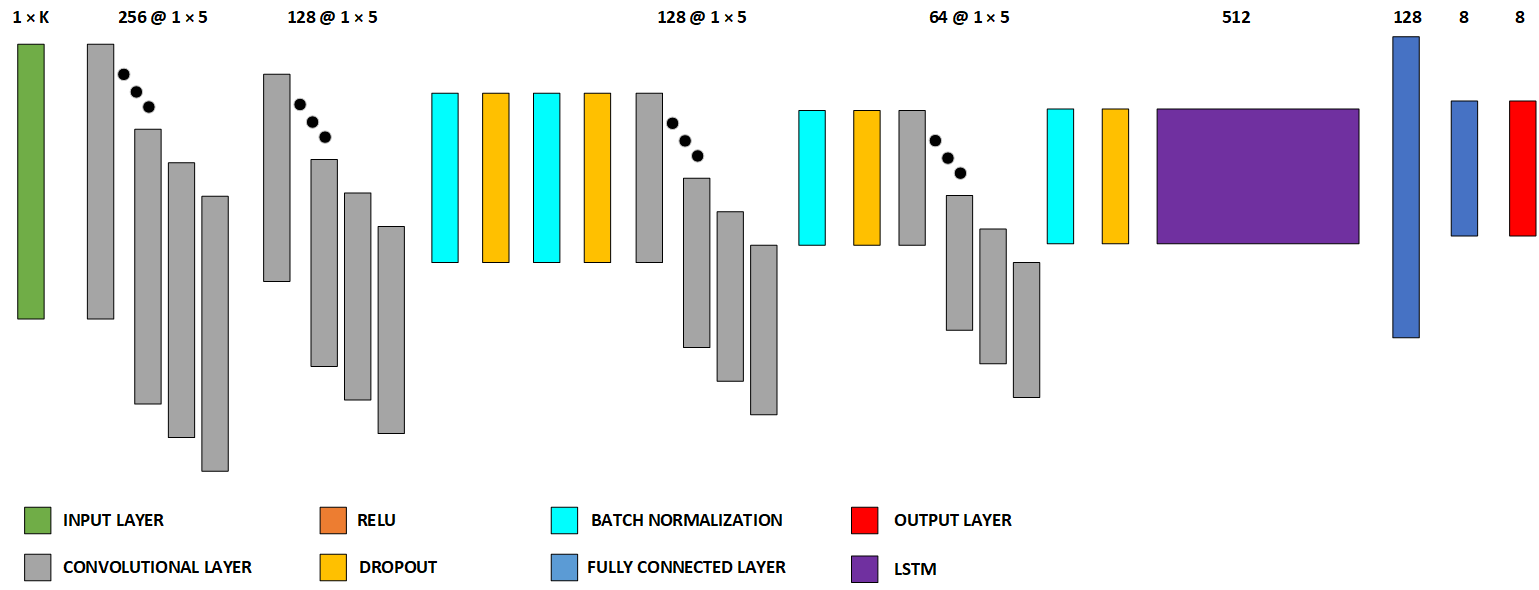}
\caption{CNN-LSTM architecture}
\label{CNN_LSTMModel}
\end{center}
\end{figure}

The CNN-LSTM method shown in Figure \ref{CNN_LSTMModel} has been applied to combine the feature extraction of CNN networks and the long term dependencies of LSTM. The first layer is consisted of 256 filters and second layer is consisted of 128 filters with the kernel size of 1 × 5 and stride 1. After that, batch normalization is carried out sending its output to the dropout layer with the ratio of 0.25. Then 2 set of convolutional layer, batch normalization and dropout is carried out. Filter sizes of this convloutional layer is 128 and 64 respectively. Next, LSTM block with 512 node is performed. It is continued 2 Fully Connected Layers (FCN) with 2000 and 8. Finally, output layer has 8 output for emotions as same as the other 2 architectures.

Using convolutional neural network architecture, deep features are fed into the three different deep and hybrid deep neural networks, namely convolutional neural network (CNN), long short-term memory network (LSTM), CNN-LSTM. Thence, speech emotion recognition procedure is implemented using deep features.

\section{Experimental results}\label{sec4}

In this study, 6 different dataset is used to demonstrate performance and generalization capability of proposed method. Details of these datasets are given in Section 3.1. All datasets are divided into 80{\%} training and 20{\%} test. The performance of the proposed methods is calculated using the accuracy formula given in (1). In (1), every class can be named as $ C_{x} $ (happy, neutral etc.). TP (True Positive) denotes audio belonging to the $ C_{x} $ is correctly classified as $ C_{x} $. FP (False Positive) represents all non-$ C_{x} $ samples classified as $ C_{x} $. TN (True Negative) is all non- $ C_{x} $ samples not classified as $ C_{x} $. FN (False Negative) represents all $ C_{x} $ samples not classified as $ C_{x} $. 

\label{equations}
\begin{equation}
Accuracy = \frac{TP+TN}{TP+TN+FP+FN}
\end{equation}

Evaluation results of machine learning based methods are given in Table \ref{table:Table2} and Table \ref{table:Table3}. Table \ref{table:Table2} shows the results where MFCC is used as the feeature extraction in machine learning-based methods, and Table \ref{table:Table3} shows the results where Mel-Spectrogram is the feature extraction. In Table \ref{table:Table2}, SVM is can be considered to be more successful as it gives the best result in 4 of 6 datasets. According to In Table \ref{table:Table3}, Random Forest has the best results in all datasets. When Table \ref{table:Table2} and Table \ref{table:Table3} are evaluated together, usage of Mel-Spectrogram and Random forest jointly gives more successful results.

Considering that machine learning-based methods did not show sufficient performance, MFCC and Mel feature extraction methods are used together with the deep learning architectures given in Section 3.4. The MFCC and Mel-spectrogram features of the raw audio signals are extracted and these features are given as an input to the deep networks. Table \ref{table:Table4} and Table \ref{table:Table5} show results using MFCC features and Mel-spectrogram features as inputs, respectively. It can be understood that, using MFCC and CNN together gives best results in all datasets. In addition, it is clearly seen that this approach is superior to machine learning-based methods.

\begin{table}[]
\caption{Evaluation results of machine learning methods with MFCC features}
\begin{tabular}{ccccccc}
\hline
\textbf{DATASET} & \textbf{SVM}   & \textbf{RF}    & \textbf{DT} & \textbf{NB} & \textbf{MV}    & \textbf{STCK} \\ \hline
EMO-DB           & \textbf{81.30} & 71.14          & 52.03       & 51.26       & 64.92          & 69.91         \\
RAVDESS          & \textbf{71.06} & 69.53          & 54.65       & 29.53       & 49.81          & 40.64         \\
TESS             & 97.23          & 98.23          & 89.2        & 86.04       & \textbf{98.52} & 95.95         \\
CREMA            & \textbf{50.26} & 45.66          & 44.50       & 30.75       & 42.65          & 41.25         \\
SAVEE            & \textbf{71.20} & 69.72          & 48.61       & 48.33       & 70.55          & 63.05         \\
TESS+RAVDESS     & 85.62          & \textbf{86.47} & 70.12       & 64.08       & 82.01          & 67.51         \\ \hline
\end{tabular}
\begin{tablenotes}
      \small
      \item The bold values show the best performance
    \end{tablenotes}
\label{table:Table2}
\end{table}

\begin{table}[]
\caption{Evaluation results of machine learning methods with Mel-Spectrogram}
\begin{tabular}{ccccccc}
\hline
\textbf{DATASET} & \textbf{SVM} & \textbf{RF}    & \textbf{DT}                  & \textbf{NB} & \textbf{MV} & \textbf{STCK} \\ \hline
EMO-DB           & 76.01        & \textbf{79.35} & 60.19                        & 39.55       & 69.65       & 59.70         \\
RAVDESS          & 27.08        & \textbf{68.98} & 45.69                        & 25.37       & 57.12       & 68.24         \\
TESS             & 84.61        & \textbf{97.47} & 89.38                        & 56.66       & 95.47       & 97.0          \\
CREMA            & 33.78        & \textbf{62.00} & {\color[HTML]{333333} 50.57} & 22.69       & 44.53       & 49.89         \\
SAVEE            & 63.05        & \textbf{75.83} & 65.55                        & 43.33       & 70.5        & 69.85         \\
TESS+RAVDESS     & 57.11        & \textbf{87.45} & 78.25                        & 38.55       & 80.81       & 87.29         \\ \hline
\end{tabular}
\begin{tablenotes}
      \small
      \item The bold values show the best performance
    \end{tablenotes}
\label{table:Table3}
\end{table}

\begin{table}[]
\caption{Evaluation results of deep learning methods with MFCC features}
\begin{tabular}{cccc}
\hline
\textbf{DATASET} & \textbf{CNN}   & \textbf{LSTM} & \textbf{CNN-LSTM} \\ \hline
EMO-DB           & \textbf{87.47} & 82.86         & 83.23             \\
RAVDESS          & \textbf{82.98} & 58.21         & 59.48             \\
TESS             & 97.40 & 97.46         & \textbf{98.57}             \\
CREMA            & \textbf{64.50} & 52.55         & 59.53             \\
SAVEE            & \textbf{83.33} & 70.48         & 70.87             \\
TESS+RAVDESS     & \textbf{94.61} & 85.53         & 87.22             \\ \hline
\end{tabular}
\begin{tablenotes}
      \small
      \item The bold values show the best performance
    \end{tablenotes}
\label{table:Table4}
\end{table}

\begin{table}[]
\caption{Evaluation results of deep learning methods with Mel-Spectrogram features}
\begin{tabular}{cccc}
\hline
\textbf{DATASET} & \textbf{CNN}   & \textbf{LSTM}  & \textbf{CNN-LSTM} \\ \hline
EMO-DB           & \textbf{86.53} & 85.66          & 58.25             \\
RAVDESS          & \textbf{57.87} & 50.57          & 27.66             \\
TESS             & 95.29          & \textbf{95.71} & 93.45             \\
CREMA            & 54.12          & \textbf{56.21} & 44.33             \\
SAVEE            & 71.87          & \textbf{75.34} & 55.55             \\
TESS+RAVDESS     & \textbf{81.48} & 80.42          & 79.95             \\ \hline
\end{tabular}
\begin{tablenotes}
      \small
      \item The bold values show the best performance
    \end{tablenotes}
\label{table:Table5}
\end{table}

Instead of extracting distinctive features of the audio signal using feature extraction methods such as MFCC or Mel-Spectrogram, the use of raw audio as input for deep learning models is analyzed. Here, the most important aim is to eliminate human intervention and to determine the most characteristic features automatically with the deep learning approach. Table \ref{table:Table6} shows end-to-end deep learning result of models given in Section 3.4. The training parameters of the deep architectures are  given in Table \ref{table:Table7}. It can be seen in Table 7, the CNN-based deep learning approach give the best results in 5 out of 6 datasets. Intriguingly, these results show that long-term dependencies features seem to play a minor role at Speech Emotion Recognition.

\begin{table}[]
\caption{Evaluation results of deep learning methods with raw audio}
\begin{tabular}{cccc}
\hline
\textbf{DATASETS} & \textbf{CNN}   & \textbf{LSTM}  & \textbf{CNN-LSTM} \\ \hline
EMO-DB            & \textbf{90.34} & 89.42          & 85.35             \\
RAVDESS           & \textbf{90.42} & 73.01          & 86.09             \\
TESS              & 99.46          & \textbf{99.48}          & 98.91             \\
CREMA             & \textbf{69.72} & 61.59          & 63.72             \\
SAVEE             & \textbf{85.76} & 82.29          & 74.61             \\
TESS+RAVDESS      & \textbf{95.86} & 91.27          & 90.50             \\ \hline
\end{tabular}
\begin{tablenotes}
      \small
      \item The bold values show the best performance
    \end{tablenotes}
\label{table:Table6}
\end{table}

\begin{figure}[htb!]
\begin{center}

    \begin{subfigure}{0.4\textwidth}
    \includegraphics[width=\linewidth]{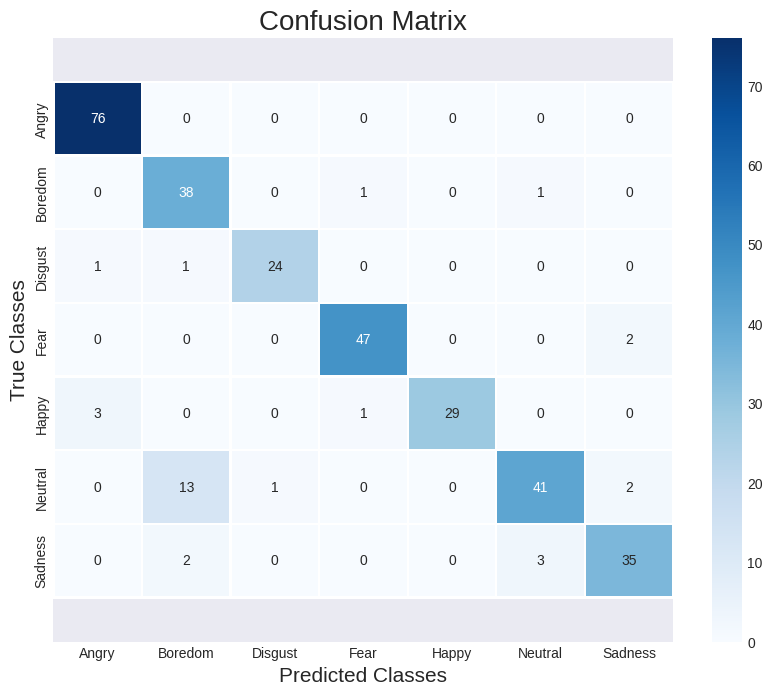}
    \caption{}
    \end{subfigure}\hfil 
    \begin{subfigure}{0.4\textwidth}
    \includegraphics[width=\linewidth]{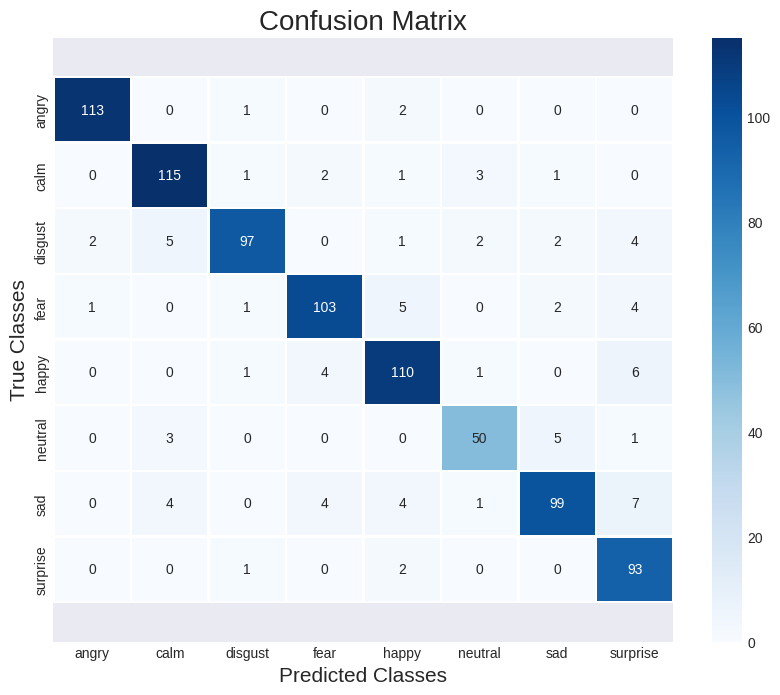}
    \caption{}
    \end{subfigure}\hfil 
    \begin{subfigure}{0.4\textwidth}
    \includegraphics[width=\linewidth]{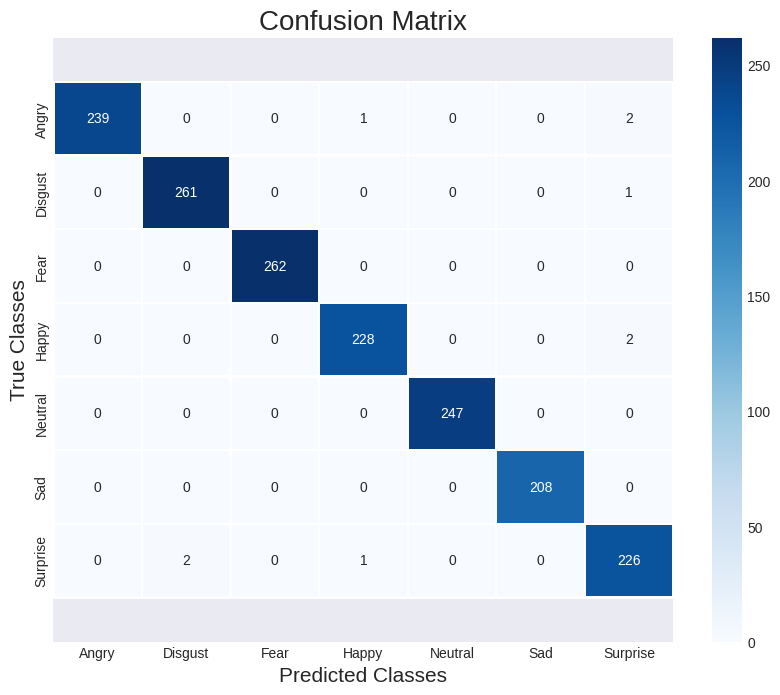}
    \caption{}
    \end{subfigure}\hfil 
    \begin{subfigure}{0.4\textwidth}
    \includegraphics[width=\linewidth]{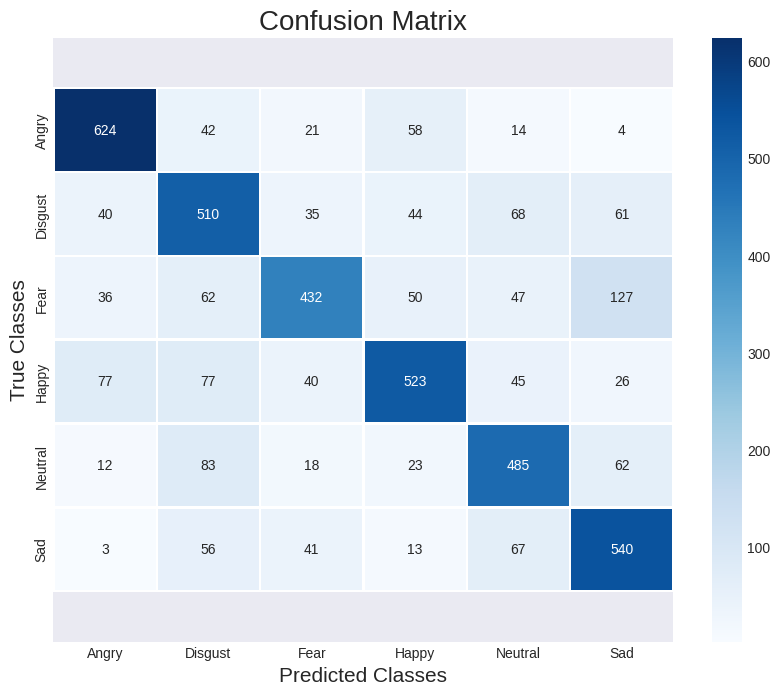}
    \caption{}
    \end{subfigure}
    \begin{subfigure}{0.4\textwidth}
    \includegraphics[width=\linewidth]{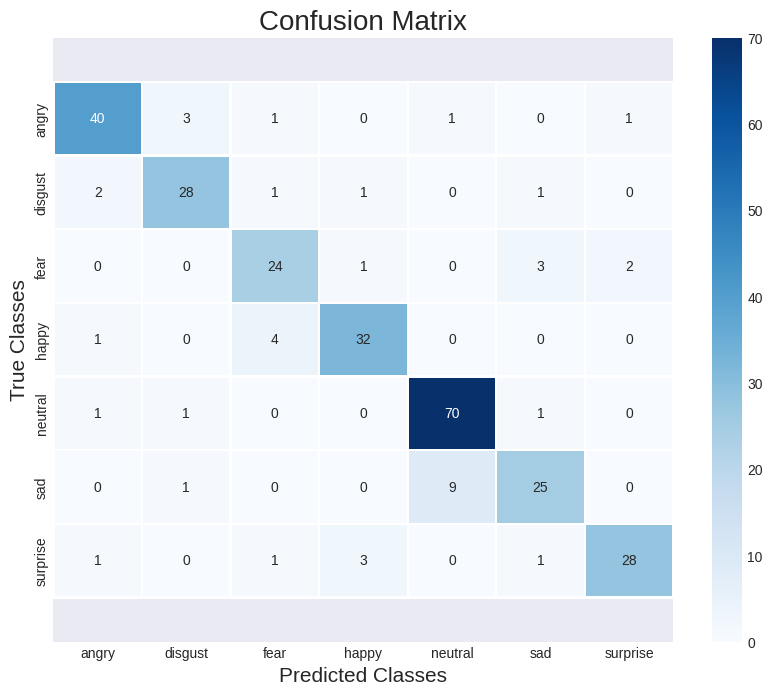}
    \caption{}
    \end{subfigure}
    \begin{subfigure}{0.4\textwidth}
    \includegraphics[width=\linewidth]{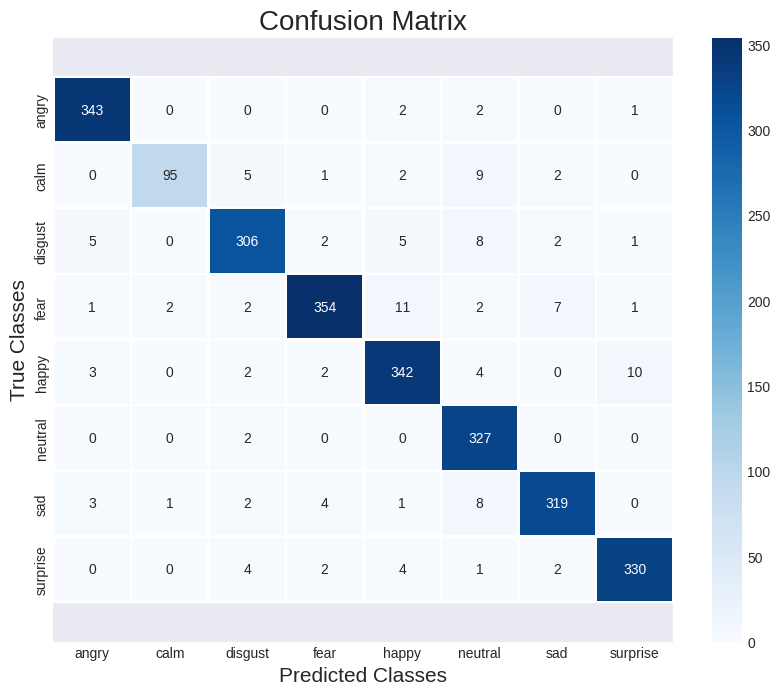}
    \caption{}
    \end{subfigure}
\caption{The confusion matrix obtained using Proposed method on various datasets (a) EMO-DB (b) RAVDESS (c) TESS (d) CREMA (e) SAVEE (f) TESS+RAVDESS.}
\label{FigConfusionMatrix}
\end{center}
\end{figure}

\begin{table}[]
\caption{Training parameters}
\begin{tabular}{cccc}
\hline
\textbf{Parameter}     & \textbf{\begin{tabular}[c]{@{}c@{}}CNN \\ Value\end{tabular}} & \textbf{\begin{tabular}[c]{@{}c@{}}LSTM \\ Value\end{tabular}} & \textbf{\begin{tabular}[c]{@{}c@{}}CNN-LSTM \\ Value\end{tabular}} \\ \hline
Learning rate          & 0.001                                                         & 0.001                                                          & 0.001                                                              \\
Epoch                  & 500                                                           & 80                                                             & 200                                                                \\
Batch                  & 8                                                             & 8                                                              & 8                                                                  \\
Optimization algorithm & ADAM                                                          & ADAM                                                           & ADAM                                                               \\ \hline
\end{tabular}
\label{table:Table7}
\end{table}

Table \ref{table:Table8} shows the comparison of the recent and proposed methods for sound emotion recognition in the literature. The comparison is carried out on the EMO-DB, RAVDESS, SAVEE, TESS, CREMA, TESS+RAVDESS datasets. The literature results used in comparison in Table \ref{table:Table8} are provided on their papers. In this table, proposed method refers CNN method with raw audio. As seen in Table \ref{table:Table8}, proposed method has superior performance results in 5 out of 6 datasets and has second best result in the other.

The confusion matrices obtained using our proposed method for various datasets with CNN are shown in Fig. \ref{FigConfusionMatrix}. The confusion matrices show how many times an observation's predicted class (the column in the table) matches its true class (the row in the table). Thus, it is possible to examine the mixing ratio between classes in detail. As seen on the confusion matrices, it is clear that the proposed method does not have a definite and continuous confusion tendency between classes.

\begin{table}[]
\caption{Comparison of proposed method and literature}
\begin{tabular}{ccccccc}
\hline
\textbf{Method} & \textbf{EMO-DB} & \textbf{RAVDESS} & \textbf{SAVEE} & \textbf{TESS} & \textbf{CREMA} & \textbf{TESS+RAVDESS} \\ \hline
\citep{Ozseven2019}          & 84.62           & -                & 72.39         & -              & -               & -                      \\
\citep{Bhavan2019}           & 92.45           & 75.69            & -             & -              & -               & -                      \\
\citep{Suganya2019}          & 85.62           & -                & -             & -              & -               & -                      \\
\citep{Assuncao2020}         & 80.40           & 71.60            & 70.40         & -              & -               & -                      \\
\citep{issa2020speech}       & 86.10           & 71.61            & -             & -              & -               & -                      \\
\citep{Nyugen2020}           & 89.00           & -                & 69.00         & -              & -               & -                      \\
\citep{Asiya2021}            & -               & -                & -             & -              & -               & \textit{89.00}          \\
\citep{Ristea2021}           & -               & -                &  -            & -              & 68.12           & -                      \\
\citep{tuncer2021automated}  & 90.09           & \textit{87.43}   & \textit{84.79}& -              & -               & -                      \\
\citep{Krishnan2021}         & -               & -                & -             & 93.3           & -               & -                      \\
\citep{Belumentals2022}      & -               & -                & -             & -              & -               & 86.02                 \\
\citep{Singh2023}            & 87.00           & -                & 80.00         & -              & -               & -                 \\
Proposed Method              & \textit{90.34}  & \textbf{90.42}   & \textbf{85.76}& \textbf{99.46} & \textbf{69.72}  & \textbf{95.86}                 \\ \hline
\end{tabular}
\begin{tablenotes}
      \small
      \item The bold values show the best performance, italic values show the second best performance
    \end{tablenotes}
\label{table:Table8}
\end{table}

\section{Conclusion}\label{sec5}

In this work, an end-to-end deep learning based approach is proposed for speech emotion recognition. In the proposed method, raw audio signals are employed as input. Three different deep learning architecture named as CNN, LSTM, and CNN-LSTM are composed. Feature extraction capability and performance of proposed architectures  are compared with MFCC and Mel-Spectrogram. Also, traditional machine learning-based approaches are combined with MFCC and Mel-Spectrogram and their performances are compared with proposed method. Experimental results show that the CNN based deep learning approach reveals remarkable results with maximum 99,46{\%} of accuracy. In future, transfer learning techniques will be used to improve performance.

\printcredits

\section*{Competing Interests}

The authors declare that they have no conflict of interest.

\bibliographystyle{cas-model2-names}

\bibliography{cas-refs}





\end{document}